\documentclass{elsart1p}
\usepackage{graphicx}
\usepackage{amssymb}
\begin{document}
\begin{frontmatter}
%
%
%
%
%
\title{Effects of running couplings on jet conversion photons}
%
%

\author{Lusaka Bhattacharya}

\address{Saha Institute of Nuclear Physics, 1/AF Bidhannagar, Kolkata,
  Pin-700064, India}

\begin{abstract}

We calculate photons from jet-plasma interaction considering 
collisional and radiative energy loss of jet parton. 
The phase space distribution of the 
participating jet is dynamically evolved by solving Fokker-Planck 
equation. We treat the strong coupling constant ($\alpha_s$) as function 
of momentum and temperature while calculating the drag and diffusion 
coefficients. It is observed that the quenching factor is 
substantially modified as compared to the case when $\alpha_s$ is 
taken as constant. It is shown that the Phenix data is reasonably well 
reproduced when contributions from all the relevant sources are 
taken into account. 

\end{abstract}

\end{frontmatter}

\section{Introduction}

Heavy ion collisions have received significant attention in recent 
years. Electromagnetic probes (photons, dileptons etc) have been 
proposed to be one of the most promising tools to characterize the 
initial state of the collisions~\cite{jpr}. Because of the very nature 
of their interactions with the constituents of the system they tend to 
leave the system almost unscattered. 
Photons are produced at various stages of the evolution process. The 
initial hard scatterings (Compton and annihilation) of partons lead to 
photon production which we call hard photons. If quark gluon plasma 
(QGP) is produced initially, there are  QGP-photons from thermal Compton 
plus annihilation processes. Photons are also produced from different 
hadronic reactions from hadronic matter either formed initially 
(no QGP scenario) or realized as a result of a phase transition from QGP. 

These apart, there exits another class of photon emission process via 
the jet conversion mechanism (jet-plasma interaction)~\cite{dks} which 
occurs when a high energy jet interacts with the medium constituents 
via annihilation and Compton processes.

In current heavy ion collision experiments, the temperature $T$ is not 
only the important scale, momentum scale, 
$k$, (of the partons) is also important. Therefore running of the 
coupling in the high momentum regime ($p \sim T$) has to be taken 
into account to calculate the cross sections and the energy-loss
processes. In this work we calculate photons from jet-plasma interaction 
taking into account running of QCD coupling and both 
collisional and radiative energy losses.

The plan of the article is as follows. We discuss the formalism in the 
next section. Results will be discussed in the section 3. Finally we 
will conclude.

\section{Formalism}

\subsection{Jet-Photon Rate}

The lowest order processes for photon emission from QGP are the 
Compton scattering 
($q ({\bar q})\,g\,\rightarrow\,q ({\bar q})\,\gamma$) and 
annihilation ($q\,{\bar q}\,\rightarrow\,g\,\gamma$) process. 
The differential photon production rate for this process is 
given by~\cite{lb1}:
\begin{eqnarray} 
E\frac{dR}{d^3p}=\frac{{\mathcal N}}{2(2\pi)^8} 
\int \Pi_{1}^3\frac{d^3p_i}{2E_i}
f_{jet}({\bf{p_1}})f_2(E_2)\delta(p_1+p_2 - p_3 - p)
|{\mathcal{M}}|^2 (1\pm f_3({E_3}))
\label{photonrate}
\end{eqnarray}
where, $|{\mathcal{M}}|^2$ represents the spin averaged matrix element 
squared for one of those processes which contributes in the photon rate 
and ${{\mathcal N}}$ is the degeneracy factor of the corresponding 
process. $f_{jet}$, $f_2$ and $f_3$ are the initial state and final 
state partons. $f_2$ and $f_3$ are the Bose-Einstein or Fermi-Dirac
distribution functions.
%
%

\subsection{Fokker - Planck Equation: Parton transverse 
momentum spectra}

In the photon production rate (from jet-plasma interaction) one of the 
collision partners is assumed to be in equilibrium and the other 
(the jet) is executing random motion in the heat bath provided by quarks 
(anti-quarks) and gluons. Furthermore, the interaction of the jet is 
dominated by small angle scattering. In such scenario the evolution 
of the jet phase space distribution is governed by Fokker-Planck 
(FP) equation where the collision integral is approximated by 
appropriately defined drag ($\eta$) and diffusion coefficients~\cite{lb6}.

The drag and diffusion coefficients are infrared singular. 
The infra-red cut-off is fixed by plasma effects, where only the 
medium part is considered, completely neglecting the vacuum 
contribution leading to ambiguity in the energy loss calculation. 
If the latter part is taken into account the strong coupling should 
be running. Thus for any consistent calculation one has 
to take into consideration this fact. In that case 
$\alpha_s=\alpha_{s}(k,T)$ ($k=\sqrt{|\omega^2-q^2|}$ in this case), 
and the above integrals must be evaluated numerically where the infra-red 
cut-off is fixed by Debye mass to be solved self-consistently:
$
{m_{D}}(T) = 4\pi\,\left(1+\frac{N_F}{6}\right)\,\alpha_s(m_D(T),T)T^2
$
Here the strong coupling which we take as 
running, i. e. $\alpha_s= \alpha_s (\sqrt{|\omega^2-q^2|},T)$. 
We chose the following parametrization of $\alpha_s$ which respects 
the perturbative ultra-violet (UV) behavior and the 3D infra-red 
(IR) point~\cite{prd75054031}:
\begin{eqnarray}
\alpha_s (k,T)= \frac{u_1 \frac{k}{T}}{1+exp(u_2\frac{k}{T}-u_3)}
+ \frac{v_1}{(1+exp(v_2 \frac{k}{T}-v_3))(ln(e+ (\frac{k}{\lambda_s})^a
+(\frac{k}{\lambda_s})^b)},
\end{eqnarray}
with $k = \sqrt{|\omega^2-q^2|}$ in this case. 
The parameters $a$, $b$ and $\lambda_s$ are given by
$a=9.07$, $b=5.90$ and $\lambda_s=0.263$ GeV. 
For the limiting behavior ($k << T$) of the coupling we choose,
%
$
u_1={\alpha^*}_{3d}(1 + exp(-u_3))
$
Here ${\alpha^*}_{3d}$ and ${\alpha^*}_s$ denote the values of the IR 
fixed point of $SU(3)$ Yang-Mills theory in $d=3$ and $d=4$ dimensions, 
respectively. The remaining four parameters ($u_2=5.47, u_3=6.01, v_2=10.13$ 
and $v_3=9.27$) fit the numerical results for pure Yang-Mills theory 
obtained from the RG equations in Ref.~\cite{Braun_Gies}.

In our calculation we have considered both collisional and 
radiative energy losses in the following manner.
\begin{eqnarray}
\eta= \eta_{\rm coll} + \eta_{\rm rad}
=\frac{1}{E}\left[\left(\frac{dE}{dx}\right)_{\rm coll}+
\left(\frac{dE}{dx}\right)_{\rm rad}\right]
\end{eqnarray}
For running $\alpha_s$, the expressions for the 
collisional and radiative energy losses can be found in \cite{lb6}. 
Having known the drag and diffusion, we solve the FP equation using 
Green's function techniques (for details see Ref. \cite {rapp}).

\subsection{Space time evolution}

In order to obtain the space-time integrated rate we first note that 
the phase space distribution function for the incoming jet in the mid 
rapidity region is given by (see Ref.~\cite{prc72} for details)
\begin{eqnarray}
f_{jet}(\vec r,\vec p,t^{\prime})|_{y=0}&=&
\frac{(2\pi)^3\mathcal{P}(|{\vec w}_r|)~t_i}{\nu_q \sqrt{{t_i}^2-{z_0}^2}}
\frac{1}{p_{T}}\frac{dN}{d^2{p_{T}}dy}(p_T,t^{\prime})\delta(z_0)
\label{jetp}
\end{eqnarray}
With this jet parton phase space distribution function one can easily 
obtain jet photon yield from eqn. (1):
\begin{eqnarray}
\frac{dN^{\gamma}}{d^2p_Tdy}&=&\int d^4x ~ \frac{dN^{\gamma}}{d^4xd^2p_Tdy} 
\nonumber\\
&=&\frac{(2\pi)^3}{\nu_q}{\int_{t_i}}^{t_c}dt^{\prime}
{\int_0}^R rdr \int d\phi\mathcal{P}(\vec {w_r})
\frac{{\mathcal{N}_i}}{16(2\pi)^7E_{\gamma}}\int 
d{\hat s}d{\hat t} |\mathcal{M}_i|^2\int {dE_1 dE_2}\nonumber\\
&\times&\frac{1}{p_{1T}}\frac{dN}{dp_{1T}^2dy}(p_{1T},t^\prime)\frac{f_{2}(E_2)
(1\pm f_3(E_3))}{\sqrt{a{E_2}^2+2bE_2+c}}
\label{last}
\end{eqnarray}

\section{Results}

\begin{figure}[ht]
\begin{center}
\includegraphics[width=12 pc,angle=-90]{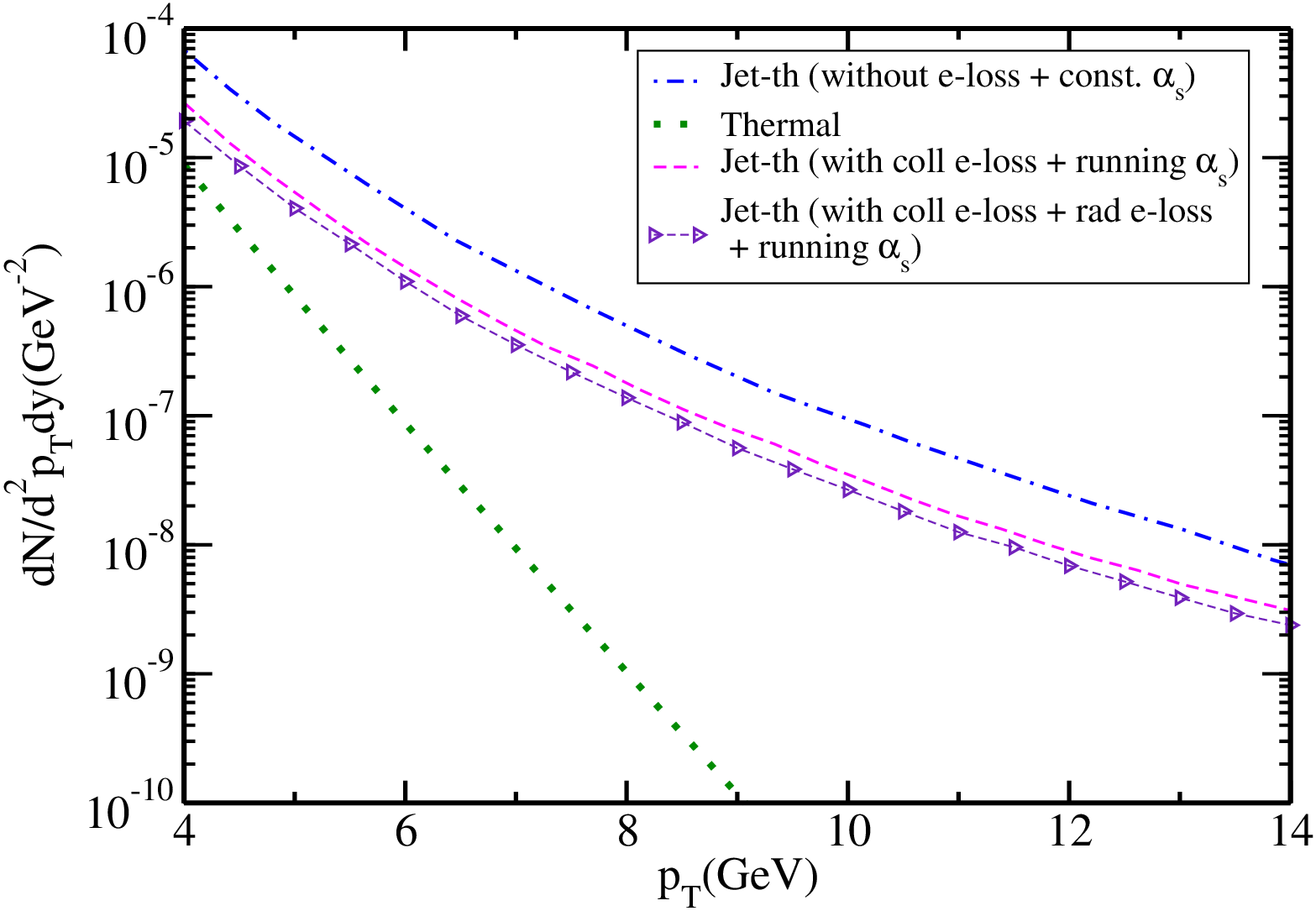}
\includegraphics[width=12 pc,angle=-90]{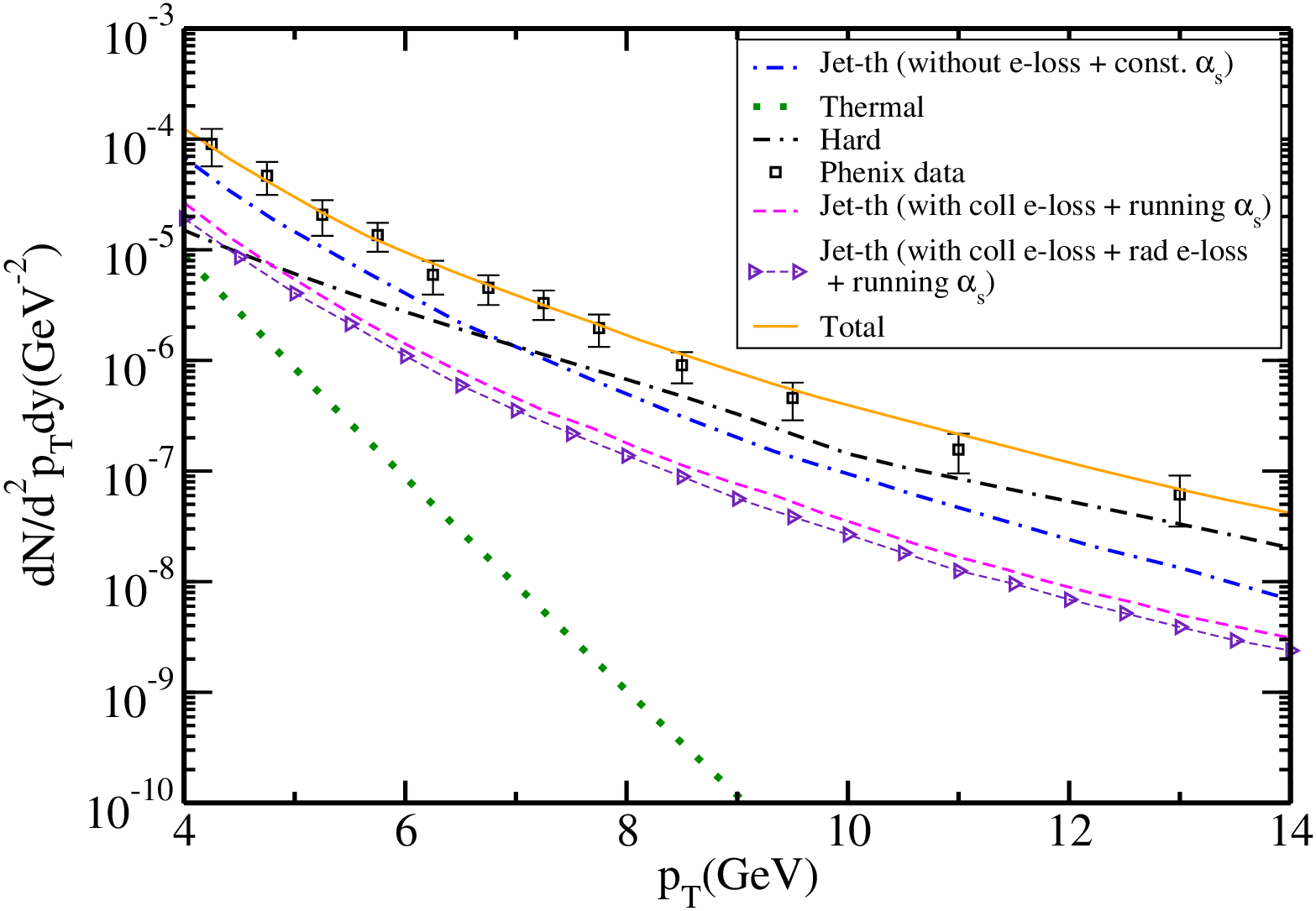}
\caption{\label{2} (color online) Left: $p_T$ distribution of 
photons at RHIC energy. 
The violet (magenta) curve denotes the photon yield from 
jet-plasma interaction with collisional (collisional + radiative) 
energy loss. The blue curve corresponds to the case without any energy 
loss and the green curve represents the thermal contribution. 
Right: Our results (the orange line which represents the total photon yield) 
is compared with the Phenix measurements of 
photon data~\protect\cite{phenix}. }
\label{fig_rhic446}
\end{center}
\end{figure}

In order to obtain the photon $p_T$ distribution we numerically integrate 
Eq. (\ref{last}). 
The results for jet-photons for RHIC energies are plotted in 
Fig.~\ref{fig_rhic446} (left) where we have taken $T_i = 446$ MeV and 
$t_i = 0.147$ fm/c.
We find that the yield                               
is decreased with the inclusion of both the energy loss
mechanisms as compared to the case when only collisional energy loss
is considered. It is to be noted that when one considers
collisional energy loss alone the yield with constant
$\alpha_s$ is more compared to the situation when running
$\alpha_s$ is taken into account (see Fig. \ref{fig_rhic446} left).

In order to compare our results with high $p_T$ photon data measured by 
the PHENIX collaboration~\cite{phenix}, we have to evaluate the 
contributions to the photons from other sources, that might contribute 
in this $p_T$ range. In Fig.~\ref{fig_rhic446} (right) the results for 
jet-photons corresponding to the RHIC energies are shown, where we have taken 
$T_i = 446$ MeV and $t_i = 0.147$ fm/c. The individual contributions 
from hard and bremsstrahlung processes~\cite{Owens} are also shown for 
comparison.

\section{Summary}

We have calculated the transverse momentum distribution of
photons from jet plasma interaction with running coupling, i. e. 
with $\alpha_s = \alpha_s(k,T)$ where we have included both collisional 
and radiative energy losses. Using running coupling we find that the 
depletion in the photon $p_T$ spectra is by a factor of $2 - 2.5$ 
more as compared to the case with constant coupling for RHIC energies
Phenix photon data have been contrasted with the present calculation 
and the data seem to have been reproduced well in the low 
$p_T$ domain. The energy of the jet quark to produce photons in this 
range ($4 < p_T < 14$) is such that collisional energy loss plays 
important role here. It is shown that inclusion of radiative energy 
loss also describes the data reasonable well.

\label{}


\begin{thebibliography}{50}

\bibitem{jpr}  J. Alam, S. Sarkar, P. Roy, T. Hatsuda, and B. Sinha,
Ann. Phys. {\bf 286} 159 (2000).

\bibitem{dks} R. J. Fries, B. Muller, and D. K. Srivastava, 
Phys. Rev. Lett. {\bf 90}, 132301 (2003).

\bibitem{lb1} L. Bhattacharya and P. Roy, Eur. Phys. J. C{\bf 69} 445 (2010). 

\bibitem{lb6} L. Bhattacharya and P. Roy, arxiv:1101.3869 [hep-ph][Accepted
  for publication in Journal of Phys. G].


\bibitem{prd75054031} J. Braun and H-J. Pirner, Phys. Rev. D{\bf 75}, 
054031 (2007).


\bibitem{Braun_Gies} J. Braun and H. Gies, J. High energy Phys. 
{\bf 06} 024 (2006).

\bibitem{rapp} H. V. Hees and R. Rapp, Phys. ReV. C{\bf 71}, 034907 (2005).

\bibitem{prc72} S. Turbide, C. Gale, S. Jeon and G. D. Moore, 
Phys. Rev. C{\bf 72}, 014906 (2005).

\bibitem{phenix}  S. S. Adler et al., Phys. Rev. Lett. {\bf 98} 012002 (2007).
\bibitem{Owens} J. F. Owens, Reviews of Modern Physics, Vol-59, No. 2, 
465 (1987).




\end{thebibliography}
\end{document}